\documentclass[reprint,aps,twoside,prd,superscriptaddress,nofootinbib,longbibliography]{revtex4-2}
\pdfoutput=0 
\usepackage{graphicx}
\usepackage[T1]{fontenc}
\usepackage{amssymb}
\usepackage{amsmath}
\usepackage{amsthm}
\usepackage{tensor,mathtools}
\usepackage{nicefrac}
\usepackage{dcolumn}
\usepackage{cancel} 
\usepackage{physics}
\usepackage{makecell}

\usepackage{enumerate}

\usepackage[version=4]{mhchem}

\usepackage[dvipsnames]{xcolor} 

\usepackage{tikz,pgfplots} 
 \pgfplotsset{compat=1.18}
\usetikzlibrary{decorations.pathreplacing} 
\usetikzlibrary{patterns}
\usetikzlibrary{calc}
\usetikzlibrary{arrows.meta,arrows}

\usepackage{booktabs}
\usepackage{pifont}
\newcommand{\cmark}{\textcolor{green!60!black}{\ding{51}}}
\newcommand{\xmark}{\textcolor{red!75!black}{\ding{55}}}

\newtheorem*{theorem*}{Theorem}

\usepackage{siunitx} 
\DeclareSIUnit\year{yr}
\DeclareSIUnit \parsec {pc}
\DeclareSIUnit{\eotvos}{E}
\usepackage{comment}

\usepackage{url}

\usepackage{aasmacros} 

\newcommand\numberthis{\addtocounter{equation}{1}\tag{\theequation}}

\usepackage[colorlinks]{hyperref}
\hypersetup{
     breaklinks=true,
    pdfstartview={FitH},    
    colorlinks=true,       
    linkcolor=blue,          
    citecolor=red,        
    filecolor=magenta,      
    urlcolor=blue,           
    anchorcolor=green,      
    linktocpage=true
}
\usepackage{cleveref}

\newcommand{\afffias}{Frankfurt Institute for Advanced Studies (FIAS), Ruth-Moufang-Str.~1, 60438 Frankfurt am Main, Germany}
\newcommand{\affgoethe}{Physics Department, Goethe University, Max-von-Laue-Str.~1, 60438 Frankfurt am Main, Germany}
\newcommand{\afftartu}{Institute of Physics, University of Tartu, W. Ostwaldi 1, 50411 Tartu, Estonia}

\date{\today}

\begin{document}

\title{Tidal Forces in the Presence of Torsion and Nonmetricity}

\author{A.~van de Venn}
\email{armin.van.de.venn@ut.ee}
\thanks{corresponding author}
\affiliation{\afftartu}
\author{M.~Netz-Marzola}
\email{marzola@fias.uni-frankfurt.de}
\affiliation{\afffias}
\affiliation{\affgoethe}
\author{D.~Vasak}
\email{vasak@fias.uni-frankfurt.de}
\affiliation{\afffias}
\author{J.~Struckmeier}
\email{struckmeier@fias.uni-frankfurt.de}
\affiliation{\afffias}

\begin{abstract}
This work investigates how torsion and nonmetricity modify tidal accelerations in metric-affine gravity. We derive a projected deviation equation that generalizes the standard geodesic deviation equation to metric-affine geometry, and apply it to the relative acceleration of neighboring autoparallels in the weak-field, nonrelativistic limit. In this regime, the tidal acceleration separates into the usual Newtonian contribution and linear post-Riemannian corrections sourced by torsion and nonmetricity. By decomposing torsion and nonmetricity into their irreducible Lorentz components, we identify the corresponding signatures in the tidal tensor and discuss to what extent these contributions can be distinguished. We then show how future direct tidal measurements could be translated into benchmark bounds on post-Riemannian tidal contributions, assuming probe dynamics sensitive to the affine connection. Our results suggest that the tidal acceleration may provide a systematic route toward probing post-Riemannian spacetime features in the future.
\end{abstract}

\maketitle

\section{Introduction}
General Relativity (GR) provides a remarkably successful description of gravitation in terms of spacetime curvature sourced by matter and energy. Nevertheless, both conceptual and phenomenological considerations motivate extensions of its geometric framework. Conceptually, the gauge principle underlying the Standard Model can also be applied to spacetime symmetries. In metric-affine gauge formulations, the coframe and linear connection are treated as independent variables associated, respectively, with translations and local linear transformations~\cite{Hehl76,Hehl95,Blagojevic12,Utiyama56,Kibble61}. Placing gravity in this common gauge-theoretic language may provide a useful starting point for attempts to unify it with the interactions of the Standard Model. Within this framework, torsion and nonmetricity arise naturally as post-Riemannian geometric quantities rather than being imposed to vanish a priori~\cite{Percacci20}.
Phenomenologically, the evidence for late-time cosmic acceleration and dark matter, together with the increasing precision of cosmological, compact-object, and gravitational-wave observations, motivates investigating whether additional geometric degrees of freedom can produce detectable effects beyond GR. In metric-affine theories, components of torsion and nonmetricity can contribute effectively to dark-energy or dark-matter phenomenology~\cite{vandeVenn22,Benisty21,delaCruz21,TARASOV23}, modify inflationary dynamics, the late-time expansion, and cosmological perturbations~\cite{Racioppi24,Zhang19,Aoki23}, and affect compact-object geometries and the gravitational-wave spectrum~\cite{Bahamonde22,Bahamonde25}. Numerous additional studies have explored the theoretical foundations, dynamical consequences, and phenomenological constraints of torsion and nonmetricity; see, for example,~\cite{Hammond02,Puetzfeld07,Trautman73,Jimenez19,Kostelecky08,Mao07,Carroll94,Lammerzahl97, Luz17, Luz19, LuzMena19, Luz25, Agarwal25, Bahamonde21, Latorre17, Iosifidis20, IosifidisKonstant23, Iosifidis22, Kirsch23, Vasak23, Struckmeier17, Ferraro18, Koide25, Koide26}.

Furthermore, variation of the matter action with respect to the independent affine connection defines the hypermomentum current. Its antisymmetric, trace, and symmetric traceless components describe the spin, dilation, and shear currents of matter, respectively. In the standard metric-affine correspondence, the spin current is associated primarily with torsion, whereas the dilation and shear currents are associated with nonmetricity. Fermionic matter, for example, sources torsion in Einstein-Cartan theory~\cite{Hehl76,Shapiro01}, while more general matter with microscopic structure may also couple to nonmetricity~\cite{Hehl95}.
Consequently, imposing the Levi-Civita connection from the outset excludes a class of possible interactions between gravitation and the internal properties of matter that cannot be described solely by the energy-momentum tensor.
Together, these considerations motivate a geometric framework in which neither torsion nor nonmetricity is required to vanish. This framework is known as metric-affine gravity (MAG).

In MAG, the connection decomposes into its Levi-Civita part, contortion, and disformation. The curvature tensor thus acquires additional contributions beyond those of Riemannian geometry. This broader geometric structure provides a unified setting encompassing Einstein-Cartan, Weyl-type, and more general post-Riemannian theories.
Although torsion and nonmetricity are auxiliary fields in some minimal Einstein-Hilbert-type models, more general invariants constructed from curvature, torsion, and nonmetricity can render components of the independent connection dynamical. Such terms arise naturally when MAG is treated as an effective field theory~\cite{Baldazzi21}. Recent investigations have therefore focused on the particle content and stability of quadratic and cubic metric-affine theories~\cite{Barker24, Bahamonde24}. Possible equivalence-principle violations associated with nonmetricity have likewise been investigated~\cite{Battista26}, while nonmetricity has also been proposed as a natural geometric manifestation of Lorentz-invariance violation~\cite{Obukhov24}. These developments make it important to identify observables that can distinguish post-Riemannian geometry from the Riemannian predictions of GR.

A particularly transparent geometric observable in any gravitational theory is the relative acceleration between neighboring worldlines, which has received comparatively limited attention in metric-affine settings. 
In GR, this relative acceleration is governed by the geodesic deviation equation, which directly relates tidal accelerations to spacetime curvature. In the weak-field, nonrelativistic limit, the equation reduces to the familiar Newtonian tidal law.
In MAG, the situation becomes more intricate. The natural generalization of the geodesic deviation equation, termed ``deviation equation'', is formulated in terms of autoparallels rather than metric geodesics \cite{vandeVenn24}. While these notions coincide in GR, they generically differ in the presence of torsion and nonmetricity. As a result, post-Riemannian degrees of freedom are expected to contribute explicitly to the deviation equation and to modify the associated tidal acceleration law.
Previous investigations of deviation and tidal phenomena have largely followed two directions: analyses within particular theories or restricted geometries---including $f(R)$ and curvature-matter-coupled theories~\cite{delaCruz13,Harko12}, teleparallel $f(T)$ and symmetric-teleparallel $f(Q)$ models~\cite{Darabi14,Beh21,Capozziello24}, Weyl-type $f(Q,T)$ gravity~\cite{Yang21}, and Riemann-Cartan geometries with torsion~\cite{Puetzfeld18,vandeVenn24}---and more general metric-affine treatments based on the unprojected deviation equation~\cite{Agashe23}. The significance of the distinction between projected and unprojected relative acceleration will be discussed later in this work. In contrast, the present analysis treats a general metric-affine geometry containing both torsion and nonmetricity simultaneously without assuming a particular gravitational theory.
In \cite{vandeVenn24}, the projected deviation equation in the presence of torsion was derived. The present work extends this analysis by systematically incorporating nonmetricity. Our aim is to obtain the fully generalized tidal force law to leading order in torsion and nonmetricity within MAG, and to analyze its physical implications.

This paper is organized as follows. In Sect.~\ref{Sect:Def_Conv} we review the general framework of metric-affine geometry and establish our conventions. Sect.~\ref{Sect:Dev_Eq} presents the derivation of the projected deviation equation in MAG. The tidal acceleration law and its weak-field, nonrelativistic limit are analyzed in Sect.~\ref{Sect:Tidal}. Section~\ref{Sect:Constr} develops a qualitative procedure for extracting possible benchmark bounds on torsion and nonmetricity from future direct tidal measurements. Finally, Sect.~\ref{Sect:Disc} summarizes the results and discusses their implications.

\section{Mathematical Framework}
\label{Sect:Def_Conv}
We consider a four-dimensional spacetime manifold equipped with a Lorentzian metric $g_{\mu\nu}$ of signature $(-,+,+,+)$ and an independent affine connection $\Gamma\indices{^\lambda_{\mu\nu}}$. No a priori assumptions are made regarding metric compatibility or symmetry of the connection. We adopt the convention $\partial_\mu \equiv \partial / \partial x^\mu$ for partial derivatives
and employ natural units in which $\hbar = c = 1$.

The affine connection is decomposed into its Levi-Civita part, contortion, and disformation according to
\begin{equation}
    \Gamma\indices{^\lambda_{\mu\nu}}
    =
    \mathring{\Gamma}\indices{^\lambda_{\mu\nu}}
    +
    K\indices{^\lambda_{\mu\nu}}
    +
    L\indices{^\lambda_{\mu\nu}},
    \label{Conn_decomp}
\end{equation}
where the Levi-Civita connection,
\begin{equation}
    \mathring{\Gamma}\indices{^\lambda_{\mu\nu}}
    \coloneqq
    \frac{1}{2} g\indices{^{\lambda\rho}}
    \left(
        \partial\indices{_\mu} g\indices{_{\nu\rho}}
        +
        \partial\indices{_\nu} g\indices{_{\mu\rho}}
        -
        \partial\indices{_\rho} g\indices{_{\mu\nu}}
    \right),
\end{equation}
is torsion-free and metric compatible. Throughout, an overcircle denotes quantities constructed from the Levi-Civita connection.

The contortion tensor $K\indices{^\lambda_{\mu\nu}}$ is defined in terms of the torsion tensor,
\begin{equation}
    S\indices{^\lambda_{\mu\nu}}
    \coloneqq
    \Gamma\indices{^\lambda_{[\mu\nu]}},
\end{equation}
as
\begin{equation}
    K\indices{^\lambda_{\mu\nu}}
    \coloneqq
    g\indices{^{\lambda\rho}}
    \left(
        S\indices{_{\mu\nu\rho}}
        +
        S\indices{_{\nu\mu\rho}}
        +
        S\indices{_{\rho\mu\nu}}
    \right).
    \label{def::contort}
\end{equation}
Geometrically, torsion characterizes the failure of infinitesimal
parallelograms, constructed by parallel transport, to close.
If two infinitesimal displacements are performed successively
along vector fields $X$ and $Y$, exchanging their order produces
a mismatch proportional to the torsion tensor.

The disformation tensor $L\indices{^\lambda_{\mu\nu}}$ is defined in terms of the nonmetricity tensor,
\begin{equation}
    Q\indices{_{\lambda\mu\nu}}
    \coloneqq
    \nabla\indices{_\lambda} g\indices{_{\mu\nu}},
\end{equation}
as
\begin{equation}
    L\indices{^\lambda_{\mu\nu}}
    \coloneqq
    \frac{1}{2}
    g\indices{^{\lambda\rho}}
    \left(
        - Q\indices{_{\mu\nu\rho}}
        - Q\indices{_{\nu\mu\rho}}
        + Q\indices{_{\rho\mu\nu}}
    \right).
\end{equation}
Nonmetricity measures the failure of the affine connection to preserve the metric under parallel transport. In a metric-compatible geometry, lengths and angles between vectors remain unchanged when transported. If $Q\indices{_{\lambda\mu\nu}} \neq 0$, the inner product of vectors varies along curves, so that lengths and angles are not preserved anymore.

The covariant derivative of a $(1,1)$-tensor $A\indices{^\alpha_\beta}$ with respect to the affine connection $\Gamma\indices{^\lambda_{\mu\nu}}$ is given by
\begin{equation}
    \nabla\indices{_\mu} A\indices{^\alpha_\beta}
    =
    \partial\indices{_\mu} A\indices{^\alpha_\beta}
    +
    \Gamma\indices{^\alpha_{\rho\mu}}
    A\indices{^\rho_\beta}
    -
    \Gamma\indices{^\rho_{\beta\mu}}
    A\indices{^\alpha_\rho}.
    \label{Cov_der_def}
\end{equation}
More generally, the covariant derivative acts on each tensor index separately, with a positive connection term for every contravariant index and a negative connection term for every covariant index, and the respective index structure in analogy to \eqref{Cov_der_def}.

The curvature tensor associated with the full connection is defined by
\begin{equation}
    R\indices{^\lambda_{\sigma\mu\nu}}
    \coloneqq
    \partial\indices{_\mu} \Gamma\indices{^\lambda_{\sigma\nu}}
    -
    \partial\indices{_\nu} \Gamma\indices{^\lambda_{\sigma\mu}}
    +
    \Gamma\indices{^\lambda_{\rho\mu}}
    \Gamma\indices{^\rho_{\sigma\nu}}
    -
    \Gamma\indices{^\lambda_{\rho\nu}}
    \Gamma\indices{^\rho_{\sigma\mu}}.
\end{equation}
Using the decomposition of the connection \eqref{Conn_decomp}, the curvature tensor can be expressed as
\begin{align}
    R\indices{^\lambda_{\sigma\mu\nu}}
    =
    &\,
    \mathring{R}\indices{^\lambda_{\sigma\mu\nu}}
    +
    2 \mathring{\nabla}\indices{_{[\mu}}
    K\indices{^\lambda_{|\sigma|\nu]}}
    +
    2 K\indices{^\lambda_{\rho[\mu}}
    K\indices{^\rho_{|\sigma|\nu]}}
    \nonumber\\
    &+
    2 \mathring{\nabla}\indices{_{[\mu}}
    L\indices{^\lambda_{|\sigma|\nu]}}
    +
    2 L\indices{^\lambda_{\rho[\mu}}
    L\indices{^\rho_{|\sigma|\nu]}}
    \nonumber\\
    &+
    2 K\indices{^\lambda_{\rho[\mu}}
    L\indices{^\rho_{|\sigma|\nu]}}
    +
    2 L\indices{^\lambda_{\rho[\mu}}
    K\indices{^\rho_{|\sigma|\nu]}}.
\end{align}
This expression makes explicit that curvature receives contributions
from three types of terms: the Riemannian curvature of the Levi-Civita
connection, derivative terms linear in torsion and nonmetricity, and
quadratic self- and mixed-interaction terms in $K$ and $L$.
Furthermore, in the presence of torsion and nonmetricity, the curvature tensor no longer
possesses the usual index symmetries of the Riemann tensor. In particular,
antisymmetry in the first pair of indices and the pair-exchange symmetry
generally do not hold.

Finally, let us discuss the distinction between geodesics and
autoparallels in a metric-affine spacetime.
A geodesic is defined as a curve that extremizes the spacetime
interval
\begin{equation}
    S = \int
    \sqrt{
        - g\indices{_{\mu\nu}}
        \frac{\mathrm{d}x\indices{^\mu}}{\mathrm{d}\lambda}
        \frac{\mathrm{d}x\indices{^\nu}}{\mathrm{d}\lambda}
    }
    \mathrm{d}\lambda,
\end{equation}
where $x\indices{^\mu}(\lambda)$ denotes the curve in local
coordinates and $\lambda$ is a curve parameter.
The condition $\delta S = 0$ yields
\begin{equation}
    \frac{\mathrm{d}^2 x\indices{^\mu}}{\mathrm{d}\lambda^2}
    +
    \mathring{\Gamma}\indices{^\mu_{\alpha\beta}}
    \frac{\mathrm{d}x\indices{^\alpha}}{\mathrm{d}\lambda}
    \frac{\mathrm{d}x\indices{^\beta}}{\mathrm{d}\lambda}
    = 0,
\end{equation}
so that only the Levi-Civita connection enters the geodesic equation.
This reflects the fact that geodesics are determined purely by
the metric structure.

An autoparallel curve, in contrast, is defined by the requirement
that its tangent vector field $X\indices{^\mu}$ be parallel
transported along itself,
\begin{equation}
    X\indices{^\alpha} \nabla\indices{_\alpha} X\indices{^\mu} = 0.
\end{equation}
In local coordinates, with
$X\indices{^\mu} = \mathrm{d}x\indices{^\mu}/\mathrm{d}\lambda$,
this condition becomes
\begin{equation}
    \frac{\mathrm{d}^2 x\indices{^\mu}}{\mathrm{d}\lambda^2}
    +
    \Gamma\indices{^\mu_{\alpha\beta}}
    \frac{\mathrm{d}x\indices{^\alpha}}{\mathrm{d}\lambda}
    \frac{\mathrm{d}x\indices{^\beta}}{\mathrm{d}\lambda}
    = 0.
    \label{autoparallel_cond}
\end{equation}
Hence, autoparallels depend on the full affine connection,
including both torsion and nonmetricity. In general, geodesics
and autoparallels coincide if and only if the symmetric part
of the affine connection reduces to the Levi-Civita connection,
i.e.\ when
\begin{equation}
    \Gamma\indices{^\mu_{(\alpha\beta)}}
    =
    \mathring{\Gamma}\indices{^\mu_{\alpha\beta}},
\end{equation}
which is equivalent to the condition
\begin{equation}
    K\indices{^\mu_{(\alpha\beta)}}
    +
    L\indices{^\mu_{(\alpha\beta)}}
    = 0.
\end{equation}

\section{Projected Deviation Equation}
\label{Sect:Dev_Eq}
In this section, we derive the projected deviation equation in a general
metric-affine spacetime with torsion and nonmetricity.
The construction closely follows our previous derivation
for torsionful but metric-compatible geometries \cite{vandeVenn24},
which we generalize here by allowing for nonmetricity.
For completeness, we briefly summarize the geometric setup,
referring to \cite{vandeVenn24} for details of the flow
construction.

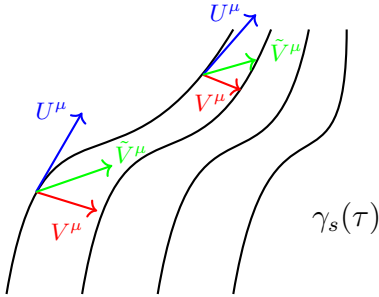
\begin{figure}[h]
	\begin{tikzpicture}
		\draw[thick] (4,-0.5) .. controls (4.5,2.5) and (5.5,0.5) .. (7,3);
        \draw[thick] (5,-0.5) .. controls (5.5,2.5) and (6.5,0.5) .. (7.5,3);
        \draw[thick] (6,-0.5) .. controls (6.5,2.5) and (7.5,0.5) .. (8,3);
        \draw[thick] (7,-0.5) .. controls (7.5,2.5) and (8.5,0.5) .. (8.5,3)
        node[xshift=0cm,yshift=-2.5cm]{\large{$\gamma_s (\tau)$}};
        \draw[thick,->,color = blue] (4.39,0.85) -- (5,1.9)
        node[xshift=-0.4cm]{$U\indices{^\mu}$};
        \draw[thick,->,color = red] (4.39,0.85) -- (5.2,0.6)
        node[xshift=-0.4cm,yshift=-0.3cm]{$V\indices{^\mu}$};
        \draw[thick,->,color = green] (4.39,0.85) -- (5.4,1.2)
        node[xshift=0.25cm,yshift=0.15cm]{$\tilde{V}\indices{^\mu}$};
        \draw[thick,->,color = blue] (6.6,2.4) -- (7.3,3.2)
        node[xshift=-0.4cm]{$U\indices{^\mu}$};
        \draw[thick,->,color = red] (6.6,2.4) -- (7.1,2.2)
        node[yshift=-0.2cm,xshift=-0.4cm]{$V\indices{^\mu}$};
        \draw[thick,->,color = green] (6.6,2.4) -- (7.3,2.6)
        node[yshift=0.1cm,xshift=0.4cm]{$\tilde{V}\indices{^\mu}$};
	\end{tikzpicture}
	\caption{The family of timelike curves $\gamma_s (\tau)$ with 
    tangent vector field $U\indices{^\mu}$ and deviation vector field $V\indices{^\mu}$. Adapted from \cite{vandeVenn24}.}
	\label{congruence}
\end{figure}

Let $U^\mu$ be a timelike vector field tangent to a congruence
of curves $\gamma_s(\tau)$,
\begin{equation}
    U^\mu = \frac{\partial \gamma_s^\mu}{\partial \tau},
\end{equation}
where $\tau$ parametrizes each curve and $s$ labels neighboring
curves within the congruence.
The connecting vector field between neighboring curves is defined as
\begin{equation}
    \tilde{V}^\mu
    =
    \frac{\partial \gamma_s^\mu}{\partial s},
\end{equation}
and represents their infinitesimal separation at a given $\tau$. Ultimately, the projected deviation equation tells us how the orthogonal separation of the curves changes along their flow.
Now since $\tilde{V}^\mu$ may in general contain components parallel
to $U^\mu$, we project onto the subspace orthogonal to $U^\mu$
using the projector
\begin{equation}
    \tensor{P}{^\mu_\nu}
    =
    \delta^\mu_\nu
    -
    \frac{U^\mu U_\nu}{U^2},
\end{equation}
where $U^2 \coloneqq U_\mu U^\mu  < 0$,
and define the deviation vector field
\begin{equation}
    V^\mu
    \coloneqq
    \tensor{P}{^\mu_\nu} \tilde{V}^\nu,
\end{equation}
which satisfies $U_\mu V^\mu = 0$, see Fig.\,\ref{congruence}.
The goal is then to compute the second covariant derivative of $V$ along $U$,
as this quantity describes the relative acceleration of nearby curves
and therefore determines the tidal dynamics.

In contrast to the metric-compatible case, the tangent vector field $U^\mu$ cannot
in general be normalized when nonmetricity is present.
To see this, consider an arbitrary vector field $X^\mu$.
The change of its squared norm along another vector field $Y^\mu$ is
\begin{equation}
    Y^\alpha \nabla_\alpha
    \left(
        g_{\mu\nu} X^\mu X^\nu
    \right)
    =
    X^\mu X^\nu Y^\alpha \nabla_\alpha g_{\mu\nu}
    +
    2 X^\mu g_{\mu\nu} Y^\alpha \nabla_\alpha X^\nu.
    \label{cov_diff_norm}
\end{equation}
In a metric-compatible geometry, $\nabla_\alpha g_{\mu\nu}=0$,
and the norm of a parallel transported vector ($Y^\alpha \nabla_\alpha X^\mu = 0$) is preserved.
However, in the presence of nonmetricity,
$\nabla_\alpha g_{\mu\nu} = Q_{\alpha\mu\nu} \neq 0$,
so that even for parallel transport
the norm evolves according to
\begin{equation}
    Y^\alpha \nabla_\alpha
    \left(
        g_{\mu\nu} X^\mu X^\nu
    \right)
    =
    X^\mu X^\nu Y^\alpha Q_{\alpha\mu\nu}.
\end{equation}
Consequently, if one were to impose a normalization condition
$X^\mu X_\mu = -1$ everywhere and simultaneously require
autoparallel transport ($X^\alpha \nabla_\alpha X^\mu = 0$), this would lead to the restriction (choosing $Y=X$)
\begin{equation}
    X^\alpha X^\mu X^\nu Q_{\alpha\mu\nu} = 0.
\end{equation}
Since we aim to keep the geometry fully general and later
indeed consider autoparallel congruences, we do not impose
a normalization condition on the tangent vector field $U^\mu$ at this point.

Now specifically, what we want to compute is the relative acceleration
\begin{equation}
    a\indices{^\mu} \coloneqq P\indices{^\mu_\lambda}U\indices{^\rho}\nabla\indices{_\rho}\left(P\indices{^\lambda_\nu}U\indices{^\alpha}\nabla
    \indices{_\alpha} V\indices{^\nu}\right).
    \label{accel_def}
\end{equation}
Even if $V$ is initially orthogonal to $U$, a covariant derivative along
$U$ can generate components parallel to $U$. To ensure that the deviation
remains orthogonal to $U$, we therefore project after each derivative.
As a preliminary step, we start by computing the derivative of the projector in the direction of $U$. We find
\begin{align*}
    U^\beta \nabla_\beta \tensor{P}{^\rho_\lambda} &= 
    -U^\beta \nabla_\beta\left[\frac{1}{U^\mu U_\mu}U^\rho U_\lambda\right]\\
    &= \frac{1}{\left(U^\mu U_\mu\right)^2}U^\beta \nabla_\beta\left(U^\mu U_\mu\right)U^\rho U_\lambda\\
    &\hphantom{=}- \frac{1}{\left(U^\mu U_\mu\right)}U^\beta \nabla_\beta\left(U^\rho U_\lambda\right). \numberthis
    \label{Cov_proj}
\end{align*}
Now use the fact that nonmetricity is present to obtain
\begin{equation}
    U^\beta \nabla_\beta\left(U^\mu U_\mu\right) = 
    U^\beta U^\mu U^\nu \tensor{Q}{_{\beta\mu\nu}}
    + 2 U_\nu \hat{A}^\nu
    \label{lemma1}
\end{equation}
and
\begin{equation}
    U^\beta \nabla_\beta\left(U^\rho U_\lambda\right) = 
    \hat{A}^\rho U_\lambda + \check{A}_\lambda U^\rho,
    \label{lemma2}
\end{equation}
where we employed the definitions
\begin{align}
    \hat{A}^\rho &\coloneqq U^\beta \nabla_\beta U^\rho\\
    \check{A}_\rho &\coloneqq U^\beta \nabla_\beta U_\rho.
\end{align}
Due to nonmetricity, these two accelerations are not related by simple index pulling with the metric anymore.
Indeed, they are related by
\begin{equation}
    \check{A}_\rho = \tensor{g}{_{\rho\alpha}}\hat{A}^\alpha +
    U^\alpha U^\beta \tensor{Q}{_{\alpha\beta\rho}},
\end{equation}
where we first used the symmetry of nonmetricity in its last two indices,
$Q_{\beta\rho\alpha}=Q_{\beta\alpha\rho}$, and then relabelled the dummy indices
$\alpha\leftrightarrow\beta$.

Notice that 
\begin{equation}
    \hat{A}^\rho = 
    \mathring{A}^\rho +
    U^\beta\,U^\alpha ( 
    K\indices{^\rho_{\alpha\beta}}
    +
    L\indices{^\rho_{\alpha\beta}})
\end{equation}
with $\mathring{A}^\rho$ being the geodesic acceleration. Hence, while $\hat{A}^\rho$ vanishes for particles moving on an autoparallel trajectory, we have to set $\hat{A}^\rho = U^\beta\,U^\alpha ( 
    K\indices{^\rho_{\alpha\beta}}
    +
    L\indices{^\rho_{\alpha\beta}})$
for geodesics.

Inserting \eqref{lemma1} and \eqref{lemma2} back into \eqref{Cov_proj}, we find 
\begin{align*}
    U^\beta \nabla_\beta \tensor{P}{^\rho_\lambda} = & \frac{1}{U^4}U^\rho U_\lambda \left[ U^\beta U^\mu U^\nu \tensor{Q}{_{\beta\mu\nu}}
    + 2 U_\nu \hat{A}^\nu\right]\\
    &- \frac{1}{U^2}\left[\hat{A}^\rho U_\lambda + \check{A}_\lambda U^\rho\right],\numberthis
    \label{Cov_Proj_res}
\end{align*}
where $U^4 \coloneqq (U^\mu U_\mu)^2$.

The term $U\indices{^\alpha}\nabla
\indices{_\alpha} V\indices{^\nu}$ in \eqref{accel_def} is computed in the following manner. 
Notice that since the Lie bracket of $U$ and $\tilde{V}$ vanishes (cf. \cite{vandeVenn24}), i.e. $[U,\tilde V]^\mu=0$, we have 
\begin{equation}
    [U,V]^\mu = -[U,\kappa U]^\mu
    = -\bigl(U^\alpha \partial_\alpha \kappa\bigr)U^\mu,
\end{equation}
where we used the decomposition 
$\tilde V^\mu = V^\mu + \kappa U^\mu$
with $\kappa\coloneqq U_\mu\tilde{V}^\mu/U^2$. Hence $[U,V]^\mu$ is parallel to $U^\mu$, and therefore
\begin{equation}
    P\indices{^\lambda_\mu} [U,V]^\mu = 0.
\end{equation}
Furthermore, 
\begin{align*}
    [U,V]^\mu &= U\indices{^\alpha}\partial
    \indices{_\alpha} V\indices{^\mu} - 
    V\indices{^\alpha}\partial
    \indices{_\alpha} U\indices{^\mu}\\
    &= U\indices{^\alpha}\nabla
    \indices{_\alpha} V\indices{^\mu} - V\indices{^\alpha}\nabla
    \indices{_\alpha} U\indices{^\mu} - 2 U\indices{^\alpha}V\indices{^\beta} S\indices{^\mu_{
    \beta\alpha}},\numberthis
\end{align*}
and contracting this whole equation with $P\indices{^\lambda_\mu}$ we find
\begin{equation}
    P\indices{^\lambda_\nu}U\indices{^\alpha}\nabla
    \indices{_\alpha} V\indices{^\nu} = P\indices{^\lambda_\nu}V\indices{^\alpha}\nabla
    \indices{_\alpha} U\indices{^\nu} + 2 P\indices{^\lambda_\nu}U\indices{^\alpha}V\indices{^\beta} S\indices{^\nu_{
    \beta\alpha}}.
    \label{cov_diff_half}
\end{equation}
Apart from the implicit appearance of nonmetricity in the connection, the projection of the first term on the rhs of \eqref{cov_diff_half} is the only change with respect to the corresponding formula in the metric-compatible case \cite{vandeVenn24}.

With \eqref{cov_diff_half}, we are halfway towards computing \eqref{accel_def}.
It remains to take the covariant derivative of \eqref{cov_diff_half} and project the result. After some algebra, using the symmetry properties of the involved tensors and using 
\begin{equation}
    \nabla\indices{_\alpha}\nabla
    \indices{_\beta} C\indices{^\mu} = \nabla\indices{_\beta}\nabla\indices{_\alpha} C\indices{^\mu} + R\indices{^\mu_{\lambda\alpha\beta}}
    C\indices{^\lambda} + 2S\indices{^\lambda_{\alpha\beta}}
    \nabla\indices{_\lambda}C\indices{^\mu},
\end{equation}
for any vector $C\indices{^\mu}$, we finally find
\begin{align*}
    a\indices{^\mu} = &P\indices{^\mu_\nu}R\indices{^\nu_{\lambda\alpha\beta}}U\indices{^\lambda}
    U\indices{^\alpha}V\indices{^\beta} - 
    \frac{4}{U^2} S\indices{^\lambda_{\beta\alpha}}U\indices{_\lambda}
    U\indices{^\alpha}V\indices{^\beta}
    P\indices{^\mu_\nu}\hat{A}\indices{^\nu}\\ &-
    \frac{1}{U^2}P\indices{^\mu_\nu}\hat{A}\indices{^\nu}
    \check{A}\indices{_\alpha}\tilde{V}\indices{^\alpha}
    + V\indices{^\alpha}P\indices{^\mu_\rho}\nabla\indices{_\alpha}\hat{A}\indices{^\rho}\\
    &+2P\indices{^\mu_\lambda}\Bigl[U\indices{^\alpha}
    V\indices{^\rho}
    U\indices{^\beta}\nabla\indices{_\beta}
    S\indices{^\lambda_{\rho\alpha}}
    + U\indices{^\alpha}S\indices{^\lambda_{\rho\alpha}}
    V\indices{^\beta}
    \nabla\indices{_\beta}U\indices{^\rho} \\
    &+ 2U\indices{^\alpha}U\indices{^\beta}
    V\indices{^\xi}S\indices{^\lambda_{\rho\alpha}}
    S\indices{^\rho_{\xi\beta}}
    + S\indices{^\lambda_{\rho\alpha}}V\indices{^\rho}
    \hat{A}\indices{^\alpha}\Bigr]\\
    &-\frac{2}{U^2}P\indices{^\mu_\nu}\hat{A}\indices{^\nu}U\indices{_\lambda}V\indices{^\alpha}
    \nabla\indices{_\alpha}U\indices{^\lambda} + \frac{\kappa}{U^2}P\indices{^\mu_\nu}\hat{A}\indices{^\nu}U\indices{_\rho}\hat{A}\indices{^\rho}\\
    &+ \frac{\kappa}{U^2}P\indices{^\mu_\nu}\hat{A}\indices{^\nu}
    U\indices{^\alpha}U\indices{^\beta}
    U\indices{^\lambda}Q\indices{_{\alpha\beta\lambda}}. \numberthis
    \label{dev_eq}
\end{align*}
This is the most general projected deviation equation in metric-affine geometry. Related deviation equations have been derived previously in metric-affine frameworks, see e.g. \cite{Kleyn04, Agashe23}. However, \cite{Kleyn04} is restricted to torsion, while \cite{Agashe23} considers the unprojected deviation equation and therefore includes, besides the transverse part, also contributions parallel to the tangent vector of the congruence. For the present purposes this distinction is crucial. Since we are interested in tidal forces in the conventional operational sense, namely relative accelerations of neighboring worldlines measured as spatial separations at equal time, the deviation equation must be projected orthogonally to the congruence, following \cite{Hawking73}. Equation \eqref{dev_eq} is precisely this projected deviation equation, and thus captures the transverse tidal acceleration in a completely general metric-affine spacetime.

As a consistency check, let us consider the metric-compatible limit.
If nonmetricity vanishes, $Q_{\alpha\mu\nu}=0$, the norm of $U^\mu$
is preserved under parallel transport, and we may freely impose
the normalization condition $U^\mu U_\mu = -1$.
In this case the distinction between $\check{A}$ and $\hat{A}$
disappears, so that $\check{A}^\mu = \hat{A}^\mu \equiv A^\mu$,
and we have the orthogonality $U^\mu A_\mu = 0$. Moreover, metric compatibility restores the antisymmetry of the
curvature tensor in its first pair of indices,
so that the projection in the first term of
\eqref{dev_eq} becomes redundant and reduces
to the identity on the orthogonal subspace.
Under these considerations, the deviation equation \eqref{dev_eq} reduces
precisely to the torsionful result derived in
\cite{vandeVenn24}:
\begin{align*}
    a\indices{^\mu} = &R\indices{^\mu_{\lambda\alpha\beta}}U\indices{^\lambda}
    U\indices{^\alpha}V\indices{^\beta} + 
    4 S\indices{^\lambda_{\beta\alpha}}U\indices{_\lambda}
    U\indices{^\alpha}V\indices{^\beta}
    A\indices{^\mu}\\ &+
    A\indices{^\mu}A\indices{_\alpha}V\indices{^\alpha}
    + V\indices{^\alpha}P\indices{^\mu_\rho}\nabla\indices{_\alpha}A\indices{^\rho}\\
    &+2P\indices{^\mu_\lambda}\Bigl[U\indices{^\alpha}
    V\indices{^\rho}
    U\indices{^\beta}\nabla\indices{_\beta}
    S\indices{^\lambda_{\rho\alpha}}
    + U\indices{^\alpha}S\indices{^\lambda_{\rho\alpha}}
    V\indices{^\beta}
    \nabla\indices{_\beta}U\indices{^\rho} \\
    &+ 2U\indices{^\alpha}U\indices{^\beta}
    V\indices{^\xi}S\indices{^\lambda_{\rho\alpha}}
    S\indices{^\rho_{\xi\beta}}
    + S\indices{^\lambda_{\rho\alpha}}V\indices{^\rho}
    A\indices{^\alpha}\Bigr]. \numberthis
\end{align*}

\section{Tidal Forces}
\label{Sect:Tidal}
This section is devoted to the analysis of the weak-field,
nonrelativistic limit of the projected deviation equation \eqref{dev_eq}.
In GR, this limit of the geodesic deviation
equation reproduces the familiar Newtonian tidal force law.
In the present metric-affine framework, we expect additional
contributions beyond the Newtonian term, arising from torsion
and nonmetricity.

We will shortly specialize to particle trajectories described by 
autoparallels of the full affine connection. 
A conceptual remark is therefore in order.
The question whether test particles follow geodesics or 
autoparallels in post-Riemannian geometries has been discussed 
extensively in the literature 
\cite{Aprea03,Lemos24,Adak11,Dereli00,Cebeci03,Puetzfeld14,Acedo15}.
In fact, in standard variational formulations, unpolarized,
structureless test matter without intrinsic spin or hypermomentum generally
follows geodesics rather than autoparallels of the
independent connection \cite{Puetzfeld14}. Direct
sensitivity to the independent connection instead requires an appropriate
matter-connection coupling, for example through matter carrying
hypermomentum.
Accordingly, we adopt the viewpoint advocated in
\cite{Obukhov21,Iosifidis23} that equations of motion should not be
postulated \emph{ad hoc}, but should follow from the conservation laws
once the coupling between matter and geometry has been specified.
Depending on this coupling, geodesic, autoparallel, or more general
motion may arise \cite{Obukhov21,Puetzfeld14}.

Nevertheless, for the purposes of the present analysis,
we make the explicit probe hypothesis that an affine-sensitive test body
follows an autoparallel of the full affine connection, so
that $\hat{A}^{\mu}=0$. This is adopted as a \emph{toy-model assumption} and
should not be interpreted as a universal equation of motion for ordinary
structureless test masses. This choice is geometrically
sensible since autoparallels constitute a distinguished class of curves in
MAG: their tangent vector is parallel transported by the full affine
connection. They therefore provide a simple and well-defined framework for
investigating the tidal response of an affine-sensitive probe. Moreover,
recent work has exhibited specific variational constructions that yield
autoparallel motion \cite{Heisenberg26,Csillag26}. This controlled
specialization allows us to determine how post-Riemannian fields enter the
tidal response while keeping the analysis tractable. More complete treatments based on a specified matter
coupling, including explicit hypermomentum effects, may produce additional or
modified contributions beyond those considered below.

We emphasize, however, that the projected deviation equation derived above does 
not rely on the autoparallel condition. The only assumption made was 
that the tangent vector $U$ be timelike. For particles following different
trajectories, one simply retains $\hat{A}^\mu \neq 0$, while the 
coordinate acceleration 
$\mathrm{d}^2 x\indices{^\mu}/\mathrm{d}\lambda^2$ appearing in 
$\hat{A}$ must be replaced by the corresponding equation of motion of 
the specific matter model under consideration. 

Now, with $\hat{A}^\mu=0$, equation \eqref{dev_eq} reduces to
\begin{align*}
    a\indices{^\mu} = &P\indices{^\mu_\nu}R\indices{^\nu_{\lambda\alpha\beta}}U\indices{^\lambda}
    U\indices{^\alpha}V\indices{^\beta}
    +2P\indices{^\mu_\lambda}\Bigl[U\indices{^\alpha}
    V\indices{^\rho}
    U\indices{^\beta}\nabla\indices{_\beta}
    S\indices{^\lambda_{\rho\alpha}}\\
    &+ U\indices{^\alpha}S\indices{^\lambda_{\rho\alpha}}
    V\indices{^\beta}
   \nabla\indices{_\beta}U\indices{^\rho}+ 2U\indices{^\alpha}U\indices{^\beta}
    V\indices{^\xi}S\indices{^\lambda_{\rho\alpha}}
    S\indices{^\rho_{\xi\beta}}\Bigr].\numberthis
    \label{dev_eq_interm_form}
\end{align*}
For the purposes of the present analysis, we treat torsion and nonmetricity as small post-Riemannian perturbations of a Riemannian background geometry. Our goal is to determine the leading-order corrections to the Newtonian tidal force law induced by these additional degrees of freedom.
Accordingly, we expand all relevant quantities to first order in torsion and nonmetricity, neglecting terms that are quadratic or higher order in these fields. In particular, we consistently discard mixed contributions involving products of torsion or nonmetricity with the metric perturbation $h_{\mu\nu}$.

We now specialize to the weak-field regime and write the metric as
\begin{equation}
    g_{\mu\nu} = \eta_{\mu\nu} + h_{\mu\nu},
\end{equation}
where $\eta_{\mu\nu}$ denotes the Minkowski metric and the perturbation satisfies
\begin{equation}
    |h_{\mu\nu}| \ll 1
\end{equation}
for all $\mu,\nu = 0,1,2,3$. Indices are raised and lowered with $\eta_{\mu\nu}$ to leading order.
Henceforth we retain
only terms linear in $h$, $S$, and $Q$. In particular, mixed contributions of
order $\mathcal{O}(S \sqrt{h})$ and $\mathcal{O}(Q \sqrt{h})$ will be neglected.
We furthermore assume nonrelativistic motion. Let $U^\mu = \mathrm{d}x^\mu/\mathrm{d}\tau$
denote the four-velocity, parametrized by proper time $\tau$. 
Parametrization by proper time implies normalization of the four-velocity. As we mentioned previously, the normalization of $U^\mu$ imposes 
nontrivial constraints on the nonmetricity tensor. We return to this point later.

The normalization
\begin{equation}
    g_{\mu\nu} U^\mu U^\nu = -1
\end{equation}
furthermore implies, in the weak-field limit and for small spatial velocities,
\begin{equation}
    U^0 = 1 + \mathcal{O}(h) + \mathcal{O}(v^2),
\end{equation}
\begin{equation}
    U^i = v^i + \mathcal{O}(v h) +\mathcal{O}(v^3),
\end{equation}
where
\begin{equation}
    v^i \coloneqq \frac{\mathrm{d}x^i}{\mathrm{d}t},
    \qquad
    v^2 = \delta_{ij} v^i v^j,
\end{equation}
and $v^2 \ll 1$. Thus, to leading order, the temporal component of the
four-velocity is unity, while the spatial components coincide with the
ordinary three-velocity.

For bounded motion in a weak gravitational field, the virial relation 
implies that the kinetic and gravitational potential energies are of the 
same order. Since the Newtonian potential scales as $\Phi \sim \mathcal{O}(h)$, 
we obtain
\begin{equation}
    v^2 \sim \mathcal{O}(h),
\end{equation}
and therefore
\begin{equation}
    v^i \sim \mathcal{O}(\sqrt{h}).
\end{equation}
In Sect.~\ref{Sect:Constr}, the benchmark bounds on torsion and nonmetricity are formulated with future direct measurements of tidal acceleration in mind. Such measurements are naturally associated with satellite-based experiments in weak gravitational fields. For these systems, the assumptions of bounded and nonrelativistic motion are well justified, so that the scaling relations above may be applied.

We now return to the projected deviation equation \eqref{dev_eq_interm_form}. Keeping leading order post-Riemannian contributions we have
\begin{equation}
    2U\indices{^\alpha}U\indices{^\beta}
    V\indices{^\xi}S\indices{^\lambda_{\rho\alpha}}
    S\indices{^\rho_{\xi\beta}} \approx 0,
\end{equation}
\begin{equation}
    U\indices{^\alpha}S\indices{^\lambda_{\rho\alpha}}
    V\indices{^\beta}
   \nabla\indices{_\beta}U\indices{^\rho}\approx U\indices{^\alpha}S\indices{^\lambda_{\rho\alpha}}
    V\indices{^\beta}
   \mathring{\nabla}\indices{_\beta}U\indices{^\rho},
\end{equation}
and 
\begin{equation}
    U\indices{^\alpha}
    V\indices{^\rho}
    U\indices{^\beta}\nabla\indices{_\beta}
    S\indices{^\lambda_{\rho\alpha}} \approx U\indices{^\alpha}
    V\indices{^\rho}
    U\indices{^\beta}\mathring{\nabla}\indices{_\beta}
    S\indices{^\lambda_{\rho\alpha}}.
\end{equation}
Furthermore, in the nonrelativistic limit one finds
\begin{equation}
    \mathring{\nabla}_\beta U^\rho 
    \sim \mathcal{O}(\sqrt{h}).
\end{equation}
However, this term appears multiplied by the torsion tensor and will  
therefore be neglected.
For the term involving the derivative of the torsion tensor we find
\begin{equation}
    2 P^\mu_{\ \lambda}
    U^\alpha V^\rho U^\beta
    \mathring{\nabla}_\beta
    S^\lambda_{\ \rho\alpha} \approx 2 V^j
    \left[
        \partial_0 S^\mu_{\ j0}
        - U^\mu \partial_0 S^0_{\ j0}
    \right].
\end{equation}
Finally, for the curvature contribution we obtain
\begin{align}
    P^\mu_{\ \nu}
    R^\nu_{\ \lambda\alpha\beta}
    U^\lambda U^\alpha V^\beta
    \approx\;&
    \mathring{R}^\mu_{\ \lambda\alpha\beta}
    U^\lambda U^\alpha V^\beta
    + 2 V^j \partial_{[0} K^\mu_{\ |0|j]}
    \nonumber\\
    &+ 2 V^j
    \left(
        \partial_{[0} L^\mu_{\ |0|j]}
        - U^\mu \partial_{[0} L^0_{\ |0|j]}
    \right),
\end{align}
where we used
\begin{equation}
    K^0_{\ 0\beta}
    = g^{0\rho} K_{\rho 0\beta}
    \approx \eta^{0\rho} K_{\rho 0\beta}
    = - K_{0 0\beta}
    = 0
\end{equation}
to leading order.

The purely Riemannian term reduces to
\begin{equation}
    \mathring{R}^\mu_{\ \lambda\alpha\beta}
    U^\lambda U^\alpha V^\beta
    \approx
    - \mathring{R}^\mu_{\ 0 j 0} V^j,
\end{equation}
with
\begin{equation}
    \mathring{R}_{i 0 j 0}
    = -\frac{1}{2}
      \frac{\partial^2 h_{00}}{\partial x^i \partial x^j}
    = \mathring{R}^i_{\ 0 j 0},
\end{equation}
where $h_{00} = -2\Phi$ in the Newtonian limit and $\Phi$ denotes the Newtonian potential.

Collecting all contributions, the temporal component of the relative 
acceleration vanishes,
\begin{equation}
    a^0 = 0,
\end{equation}
while the spatial components read
\begin{equation}
    a^i
    =
    - V^j \frac{\partial^2 \Phi}{\partial x^i \partial x^j}
    + 2 V^j
    \left(
        \partial_{[0} K^i_{\ |0|j]}
        + \partial_{[0} L^i_{\ |0|j]}
        + \partial_0 S^i_{\ j0}
    \right)
    \label{tidal_acceleration_law}
\end{equation}
for $i=1,2,3$. The first term on the rhs is the usual Newtonian tidal term.
This constitutes the generalized tidal force law to leading order in 
torsion and nonmetricity. At this order, the perturbations do not mix, 
so their effects may be analyzed separately.

We emphasize that the present analysis is purely qualitative: we isolate 
the leading contributions of torsion and nonmetricity to the tidal 
acceleration without committing to any assumption about their absolute 
magnitude. Any quantitative hierarchy between the metric perturbation and 
the post-Riemannian fields must arise from a specific dynamical model for 
the torsion and nonmetricity degrees of freedom. In particular, these 
fields need not scale as $\mathcal{O}(h)$.

Finally we return to the implications of the normalization of the four-vector.
Indeed, from \eqref{cov_diff_norm}, applied to $X = Y = U$, and using
\begin{equation}
    g_{\mu\nu} U^\mu U^\nu = -1
    \qquad
    \hat{A}^\mu = 0,
\end{equation}
we obtain the consistency condition
\begin{equation}
    U^\alpha U^\mu U^\nu Q_{\alpha\mu\nu} = 0.
\end{equation}
This is not an independent field equation for nonmetricity, but a
kinematical compatibility condition resulting from the simultaneous
assumptions that $U^\mu$ is parametrized by proper
time $\tau$ (equivalently ${g_{\mu\nu}U^\mu U^\nu=-1}$) and that $\tau$ is also an
affine parameter of the full-connection autoparallel. At the fully
covariant level,
$Q_{\alpha\mu\nu}U^\alpha U^\mu U^\nu=0$ constrains the combined
background-trajectory configuration. For a prescribed congruence, it
restricts the admissible nonmetricity projected along its trajectories.
For a prescribed nonmetricity background, it selects those autoparallel
congruences whose affine parameter can coincide with proper time.
In the weak-field, nonrelativistic limit considered here, the tangent
vector is fixed at leading order by $U^\mu\simeq(1,0,0,0)$. The condition
above therefore becomes the algebraic restriction
\begin{equation}
Q_{000}=0
\end{equation}
on the nonmetricity background evaluated along the reference congruence.

In a theory in which nonmetricity is determined dynamically, our analysis
applies only to background solutions and congruences for which this
condition holds along the trajectory. Whether it can be imposed
consistently on the initial data and preserved during the subsequent
evolution must be determined from the field equations. If the condition is not satisfied, an affine
parameter of the full-connection autoparallel cannot simultaneously be
proper time: one must either retain the affine parameter and allow
the tangent norm to evolve, or parametrize the curve by proper time, in which case it
satisfies the non-affinely parametrized autoparallel equation $U^\alpha\nabla_\alpha U^\mu=\zeta U^\mu$, where $\zeta = Q_{\alpha\mu\nu}U^\alpha U^\mu U^\nu/2$.

In the following, torsion and nonmetricity are decomposed into their 
irreducible representations under the Lorentz group, and the tidal 
effects associated with each sector are examined separately.

\subsection{Torsion}
The torsion tensor can be decomposed into its irreducible components 
with respect to the Lorentz group as
\begin{equation}
    S\indices{^\lambda_{\mu\nu}} = {S^{(a)}}\indices{^\lambda_{\mu\nu}} + {S^{(v)}}\indices{^\lambda_{\mu\nu}} + {S^{(t)}}\indices{^\lambda_{\mu\nu}},
\end{equation}
corresponding to the totally antisymmetric (axial), vectorial, 
and purely tensorial parts.

\begin{figure}
    \centering
    \begin{tikzpicture}[scale=2, rotate=45]

        \coordinate (A) at (0,0);      
        \coordinate (B) at (2,0);      

        \draw[dashed, gray]
            ($(A)+(-0.5,-0.9)$)
            .. controls ($(A)+(0.05,-0.4)$) and ($(A)+(0.3,0.6)$)
            .. ($(A)+(-0.5,1.2)$);

        \draw[dashed, gray]
            ($(B)+(0.1,-0.9)$)
            .. controls ($(B)+(-0.15,-0.2)$) and ($(B)+(0.15,0.6)$)
            .. ($(B)+(0,1.2)$);

        \filldraw[black] (A) circle (1pt) node[below left] {1};
        \filldraw[black] (B) circle (1pt) node[right] {2};

        \draw[-Stealth, thick, blue] (A) -- (B) node[midway, below] {$\vec{V}$};

        \draw[-Stealth, thick, green] (A) -- ++(0.5,1) node[above] {$\partial_0 \vec{s}$};

        \draw[red] (A) circle (0.07);
        \node[red] at ($(A) + (0.22, -0.4)$) {$\vec{V} \times \partial_0 \vec{s}$};

    \end{tikzpicture}
    \caption{
    Schematic illustration of the rotational contribution to the relative acceleration 
    induced by axial torsion. The dashed curves represent the 
    curves of two neighboring particles, separated by $\vec{V}$. 
    A time variation of the axial torsion vector $\partial_0 \vec{s}$ 
    generates an acceleration orthogonal to $\vec{V}$, resulting in a 
    relative rotational distortion of the congruence.
    }
    \label{Tidal_force_axial}
\end{figure}
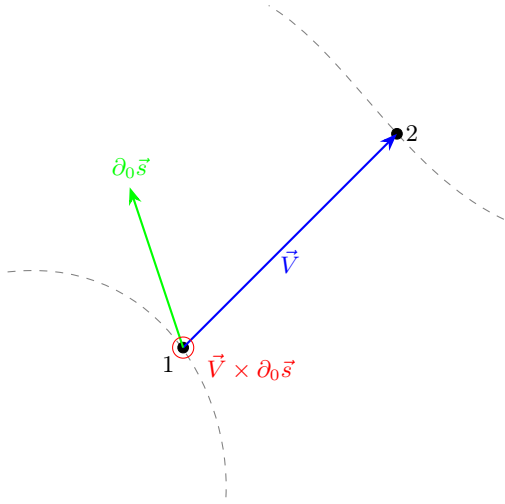

The axial torsion is given by
\begin{equation}
    {S^{(a)}}\indices{_{\lambda\mu\nu}} = \epsilon\indices{_{\lambda\mu\nu\alpha}}s\indices{^\alpha}
\end{equation}
where $\epsilon\indices{_{\lambda\mu\nu\alpha}}$ denotes the Levi-Civita tensor
and $s^\alpha$ is the axial torsion vector. However, we have $\sqrt{-g}=1+\mathcal{O}(h)$. Therefore, to leading order,
the Levi-Civita tensor coincides with the Levi-Civita symbol. In the weak-field, nonrelativistic limit, the contributions of ${S^{(a)}}$ to the tidal 
acceleration read
\begin{align*}
    \partial_0 {S^{(a)}}^i_{\ j0} &= \tensor{\eta}{^{i\alpha}} \tensor{\epsilon}{_{\alpha j 0 \mu}}\partial_0 s^\mu\\
    &= \tensor{\epsilon}{_{0ijk}}\partial_0 s^k\numberthis
\end{align*}
and
\begin{align*}
    \partial_{[0} {K^{(a)}}^i_{\ |0|j]} &= \frac{1}{2}\tensor{\eta}{^{i\rho}}\partial_0 \tensor{{S^{(a)}}}{_{\rho 0 j}}\\
    &= - \frac{1}{2}\tensor{\epsilon}{_{0ijk}}\partial_0 s^k, \numberthis
\end{align*}
where \eqref{def::contort} and total antisymmetry of torsion were used.
Together, and including the factor $2V^j$, we get 
\begin{equation}
    V^j \epsilon\indices{_{0ijk}} \partial_0 s^k,
\end{equation}
or, in three-vector notation,
\begin{equation}
    \vec V \times \partial_0 \vec s,
\end{equation}
cf. Fig.\,\ref{Tidal_force_axial}.
Since torsion carries dimension $T^{-1}$ (in units $c=1$), the term 
$\vec V \times \partial_0 \vec s$ has the correct dimension of an acceleration.
Only the components of $\partial_0 \vec s$ orthogonal to the separation 
vector $\vec V$ contribute. Consequently, the induced acceleration is 
perpendicular to $\vec V$, generating a purely rotational contribution 
to the relative motion of neighboring worldlines. 
The axial torsion therefore acts as a local ``twist'' of spacetime, 
rather than stretching or shear.

The vectorial part of torsion is given by
\begin{equation}
    {S^{(v)}}\indices{_{\lambda\mu\nu}} = -\frac{2}{3}g\indices{_{\lambda [\mu}}S\indices{_{\nu]}}\approx -\frac{2}{3}\eta\indices{_{\lambda [\mu}}S\indices{_{\nu]}},
\end{equation}
where $S\indices{_{\lambda}}\coloneqq S\indices{^\alpha_{\lambda\alpha}}$. Its contribution to the tidal acceleration reads 
\begin{equation}
    -\frac{2}{3} V^j \partial_j S^i,
\end{equation}
or equivalently,
\begin{equation}
    -\frac{2}{3} (\vec V \cdot \vec \nabla)\vec S.
\end{equation}
This term is governed by the 
directional derivative of the torsion trace along the separation 
direction. The resulting acceleration depends on how the components 
$S^i$ vary spatially and is therefore not fixed relative to 
$\vec V$. However, the symmetric part of the vectorial component $\partial_{(j}S_{i)}$ induces a 
gradient-driven tidal effect that  is structurally analogous to the 
Newtonian tidal term, but sourced by spatial variations of the 
torsion trace rather than by gradients of the gravitational potential.

The remaining tensorial part ${S^{(t)}}\indices{^\lambda_{\mu\nu}}$ is defined by 
vanishing trace while not being totally antisymmetric. 
It represents the genuinely tensorial degrees of freedom of torsion, 
which cannot be reduced to a vectorial description. 
Unlike the axial and trace components, this sector does not admit a simple 
vector interpretation in the weak-field limit, and its contribution to the 
tidal force law must therefore be retained in full index form.

Collecting all torsional contributions together with the metric 
perturbation term, the generalized tidal acceleration becomes
\begin{align*}
    a\indices{^i} = &- V^j \frac{\partial^2 \Phi}{\partial x^i \partial x^j} + 
    V\indices{^j}\epsilon\indices{_{0ijk}}
    \partial\indices{_0}s\indices{^k} 
    -\frac{2}{3} V\indices{^j}\partial\indices{_j}S\indices{^i}\\
    &+ 2V\indices{^j}\left(\partial\indices{_{[0}}{K^{(t)}}\indices{^i_{|0|j]}}
    + \partial\indices{_{0}}{S^{(t)}}\indices{^i_{j0}}\right), \numberthis
\end{align*}
where $K^{(t)}$ refers to the contortion tensor built from the purely tensorial torsion part.

Splitting the vectorial torsion contribution into its symmetric and
antisymmetric parts, the tidal acceleration may
be recast into (omitting the tensorial representation):
\begin{equation}
    a^i
    = - V^j \partial_{(j}\!\left[\partial_{i)}\Phi + \frac{2}{3}S_{i)}\right]
    + \left[\vec{V}\times\left(\partial_0 \vec{s} + \frac{1}{3}\vec{\nabla}\times\vec{S}\right)\right]^i.
    \label{tors_tidal_accel}
\end{equation}
This form makes clear that axial torsion contributes purely
rotationally, whereas vectorial torsion in general contributes to all
three types of tidal responses, namely rotation, shear, and isotropic
stretching or compression.
This finding is summarized in Table~\ref{tab:tidal_signatures}.
\begin{table}[t]
    \centering
    \setlength{\tabcolsep}{21pt}
    \begin{tabular}{l|c|c|c}
        \hline\hline
        & \textbf{$\Phi$} & \textbf{$s$} & \textbf{$S$} \\
        \hline\hline
        \makecell[l]{Stretching/\\Compression} & \cmark & \xmark & \cmark \\
        \hline
        Shear    & \cmark & \xmark & \cmark \\
        \hline
        Rotation & \xmark & \cmark & \cmark \\
        \hline\hline
    \end{tabular}
    \caption{Qualitative tidal signatures of the Newtonian $\Phi$, axial-torsion $s$, and vectorial-torsion $S$ sectors.
    Note that in vacuum, the Newtonian contribution does not possess isotropic stretching/compression since
    $\partial^i \partial_i \Phi = \Delta \Phi = 0$.}
    \label{tab:tidal_signatures}
\end{table}

\subsection{Nonmetricity}
The nonmetricity tensor can be decomposed into its irreducible components as \cite{McCrea92}
\begin{equation}
    Q\indices{_{\lambda\mu\nu}} = {Q^{(W)}}\indices{_{\lambda\mu\nu}} + {Q^{(t)}}\indices{_{\lambda\mu\nu}} + {Q^{(p)}}\indices{_{\lambda\mu\nu}} +
    {Q^{(r)}}\indices{_{\lambda\mu\nu}}.
\end{equation}
The Weyl part is given by
\begin{equation}
    {Q^{(W)}}\indices{_{\lambda\mu\nu}} = g\indices{_{\mu\nu}}W\indices{_\lambda}\approx \eta\indices{_{\mu\nu}}W\indices{_\lambda},
\end{equation}
where $W\indices{_\lambda} \coloneqq (1/4) Q\indices{_{\lambda\alpha}^\alpha}$ denotes the Weyl vector.
Geometrically, this sector describes isotropic changes of length under
parallel transport: angles are preserved, while norms are rescaled.
Its contribution to the tidal force law is
\begin{equation}
    \frac{1}{2} V\indices{^j}\partial\indices{_j}W\indices{^i} - \frac{1}{2} V\indices{^i}\partial\indices{_0}W\indices{_0},
\end{equation}
or, in three-vector notation,
\begin{equation}
    \frac{1}{2}\left[\left(\vec{V}\cdot\vec{\nabla}\right)\vec{W} - \vec{V}\partial\indices{_0}W\indices{_0}\right].
\end{equation}
The first term reflects spatial variations of the Weyl vector along the
separation direction and therefore takes the form of a directional
derivative, much like the contribution from vectorial torsion discussed
above. The second term originates from temporal variations of the time
component $W_0$ and, being proportional to $\vec V$ itself, gives rise to
a relative acceleration parallel to the separation vector, corresponding
to a longitudinal expansion or contraction.

The term ${Q^{(t)}}$ is the additional trace part of nonmetricity, given by
\begin{align*}
    {Q^{(t)}}\indices{_{\lambda\mu\nu}} &= g\indices{_{\lambda(\mu}}\Lambda\indices{_{\nu)}} - \frac{1}{4}g\indices{_{\mu\nu}}\Lambda\indices{_{\lambda}}\\
    &\approx \eta\indices{_{\lambda(\mu}}\Lambda\indices{_{\nu)}} - \frac{1}{4}\eta\indices{_{\mu\nu}}\Lambda\indices{_{\lambda}}\numberthis
\end{align*}
where $\Lambda\indices{_\mu} \coloneqq (4/9) (Q\indices{^\alpha_{\mu\alpha}} - W\indices{_\mu})$. This sector is traceful but distinct from the purely Weyl part. 
Whereas
the Weyl vector describes isotropic changes of length under parallel
transport, the $\Lambda_\mu$ sector gives rise to direction-dependent distortions of the
metric. Its contribution to the tidal force law is 
\begin{equation}
    - \frac{5}{8} V\indices{^j}\partial\indices{_j}\Lambda\indices{^i} + 
    \frac{1}{8} V\indices{^i}\partial\indices{_0}\Lambda\indices{_0},
\end{equation}
or, in three-vector notation,
\begin{equation}
    \frac{1}{8}\left[ 
    - 5\left(\vec{V}\cdot\vec{\nabla}
    \right)
    \vec{\Lambda} + \vec{V}\partial
    \indices{_0}
    \Lambda\indices{_0}
    \right].
\end{equation}
The structure of these terms closely parallels that of the Weyl sector. 
At the level of the tidal force law, the two trace sectors are therefore
algebraically analogous. Their distinction lies solely in their
geometric origin and interpretation.

The term ${Q^{(p)}}$ includes the pseudotensorial component $\Omega$ of nonmetricity. It is given by
\begin{equation}
    {Q^{(p)}}\indices{_{\lambda\mu\nu}} = \frac{1}{3}\epsilon\indices{_{\lambda\rho\sigma(\mu}}\Omega\indices{_{\nu)}^{\rho\sigma}},
\end{equation}
where $\epsilon$ denotes the Levi-Civita symbol and
\begin{equation}
    \Omega\indices{_{\lambda}^{\mu\nu}}\coloneqq
    -\left[\epsilon\indices{^{\mu\nu\rho\sigma}}Q\indices{_{\rho\sigma\lambda}} + \epsilon\indices{^{\mu\nu\rho}_\lambda}\left(\frac{3}{4}\Lambda\indices{_\rho} - W\indices{_\rho}\right)\right].
\end{equation}
Its contribution to the tidal force law reads
\begin{equation}
    V\indices{^j}\left[-\frac{1}{6}\epsilon\indices{_{ikl}}\partial\indices{_0}\Omega\indices{_{j}^{kl}} + \frac{1}{3}\epsilon\indices{_{ijk}}\partial\indices{_0}\Omega\indices{_{0}^{0k}} + \frac{1}{3}\epsilon\indices{_{ikl}}\partial\indices{_j}\Omega\indices{_{0}^{kl}}\right],
\end{equation}
where we used $\epsilon\indices{_{0ikl}}=\epsilon\indices{_{ikl}}$.
In order to make its geometric content more transparent, we introduce
\begin{align*}
    b\indices{^k}&\coloneqq\frac{1}{3} \Omega\indices{_{0}^{0k}}, 
    \qquad 
    d\indices{^i} = d\indices{_i} \coloneqq 
    \frac{1}{2} \epsilon\indices{_{ikl}}\Omega\indices{_{0}^{kl}},\\
    \Pi\indices{_{ij}}&\coloneqq 
    \frac{1}{2} \epsilon\indices{_{ikl}}\Omega\indices{_{j}^{kl}}.\numberthis
    \label{defs:odd_nonm_parts}
\end{align*}
Here $b^i$ acquires no sign change under parity transformation and is therefore parity
even, whereas $d^i$ transforms as an ordinary spatial vector and is
therefore parity odd. The object $\Pi_{ij}$ transforms as an ordinary
spatial rank-2 tensor and is likewise parity even.
In terms of these quantities, the contribution to the tidal 
acceleration can be written in vector-matrix notation as
\begin{equation}
    -\frac{1}{3}\left(\partial\indices{_0}\Pi\right)\cdot\vec{V} + 
    \vec{V}\times \partial\indices{_0}\vec{b} + \frac{2}{3}\left(\vec{V}\cdot \vec{\nabla}\right)\vec{d}.
\end{equation}
Despite the different parity assignments of $b^i$, $d^i$, and
$\Pi_{ij}$, the full expression transforms as an ordinary spatial
vector, and is therefore parity odd, as required for a spatial
acceleration.
The first term describes the action of the time-varying tensor
$\Pi\indices{_{ij}}$ on the separation vector and therefore represents a
linear distortion of the relative acceleration. The second
term, involving the cross product, has a
manifestly transverse and rotational structure, analogous to the axial
torsion contribution discussed above. Finally, the term $(\vec V\cdot\vec\nabla)\vec d$ reflects the variation
of $\vec d$ along the separation direction and thus takes the form of a
directional derivative, much like the corresponding gradient terms in the
vectorial torsion, Weyl nonmetricity, and $\Lambda_\mu$ trace
nonmetricity sectors. Unlike those contributions, however, the present
term originates from the pseudotensorial irreducible sector of nonmetricity
and therefore reflects the mirror-sensitive part of the underlying
geometry.

Finally, ${Q^{(r)}}$ denotes the remaining irreducible component of 
nonmetricity, defined by the vanishing of all traces and of the 
parity-odd contribution. It represents the genuinely tensorial 
sector that cannot be reduced to vectorial or pseudovectorial 
structures. Accordingly, its contribution to the tidal force law 
does not admit a simplification in terms of directional derivatives 
or cross-product expressions and must be retained in full index form.

Collecting all nonmetricity contributions together with the metric 
perturbation term, the generalized tidal acceleration becomes
\begin{align*}
    a\indices{^i} = &- V^j \frac{\partial^2 \Phi}{\partial x^i \partial x^j} + 
    \frac{1}{2}V\indices{^i}\left(\frac{1}{4}\partial\indices{_0}\Lambda\indices{_0} - \partial\indices{_0}W\indices{_0}\right)\\ &+ \frac{1}{2}V\indices{^j}\partial\indices{_j}\left(W\indices{^i} - \frac{5}{4}\Lambda\indices{^i} + \frac{4}{3}d\indices{^i}\right)\\
    &+ V\indices{^j}\left(\epsilon\indices{_{ijk}}\partial\indices{_0}b\indices{^k} - \frac{1}{3}\partial\indices{_0}\Pi\indices{_{ij}}\right)\\
    &+ 2V\indices{^j}\partial\indices{_{[0}}{L^{(r)}}\indices{^i_{|0|j]}}, \numberthis
\end{align*}
where $L^{(r)}$ refers to the disformation tensor built from the remainder nonmetricity part.

The tidal acceleration above may be written as (omitting the remainder component):
\begin{align*}
    a^i
    = &- V^j \Bigl[\partial_{j}\partial_{i}\Phi - \frac{1}{2}\partial_{(j}W_{i)} + \frac{5}{8}\partial_{(j}\Lambda_{i)} - \frac{2}{3}\partial_{(j}d_{i)}\\
    &-\frac{1}{8}\delta_{ij}\partial_0 \Lambda_0 + \frac{1}{2}\delta_{ij}\partial_0 W_0 + \frac{1}{3} \partial_0 \tensor{\Pi}{_{ij}}\Bigr]\\
    &+ \left[\vec{V}\times\left(\partial_0 \vec{b} + \vec{\nabla}\times\left\{-\frac{1}{4}\vec{W} + \frac{5}{16}\vec{\Lambda} - \frac{1}{3}\vec{d}\right\}\right)\right]^i. \numberthis
    \label{nonmetr_tidal_accel}
\end{align*}
We see that the irreducible sectors of nonmetricity
generally contribute to all three types of tidal
responses, namely isotropic stretching/compression, shear, and rotation; see
Table~\ref{tab:tidal_signatures_nonm}.

\begin{table}[t]
    \centering
    \setlength{\tabcolsep}{16pt}
    \begin{tabular}{l|c|c|c|c}
        \hline\hline
        & \textbf{$\Phi$} & \textbf{$W$} & \textbf{$\Lambda$} & \textbf{$\Omega$}\\
        \hline\hline
        \makecell[l]{Stretching/\\Compression} & \cmark & \cmark & \cmark & \cmark\\
        \hline
        Shear    & \cmark & \cmark & \cmark & \cmark\\
        \hline
        Rotation & \xmark & \cmark & \cmark & \cmark\\
        \hline\hline
    \end{tabular}
    \caption{Qualitative tidal signatures of the Newtonian $\Phi$, Weyl-nonmetricity $W$, trace-nonmetricity $\Lambda$, and pseudotensorial nonmetricity $\Omega$ sectors.
    Note that in vacuum, the Newtonian contribution does not possess isotropic stretching/compression since
    $\partial^i \partial_i \Phi = \Delta \Phi = 0$.}
    \label{tab:tidal_signatures_nonm}
\end{table}

In closing this section, we recall the constraint
$Q\indices{_{000}}=0$,
which, in terms of the irreducible decomposition of the nonmetricity tensor, becomes
\begin{equation}
    -W_0-\frac{3}{4}\Lambda_0+{Q^{(r)}}\indices{_{000}}=0.
    \label{kin_consist_irrep}
\end{equation}
Thus, when the Weyl, $\Lambda$-, and remainder sectors coexist,
\eqref{kin_consist_irrep} correlates the components $W_0$, $\Lambda_0$, and
$Q^{(r)}{}_{000}$, so they cannot be chosen independently.
For example, if $Q^{(r)}{}_{000}=0$, it follows that
$W_0=-3\Lambda_0/4$. The condition becomes more restrictive
in a strict single-field assumption, by which we mean that
all irreducible sectors except the one under consideration
vanish. A Weyl-only configuration then requires $W_0=0$, a
$\Lambda$-only configuration requires $\Lambda_0=0$, and a
remainder-only configuration requires $Q^{(r)}{}_{000}=0$,
along the reference congruence. Such an assumption is central for the following section and will be discussed below.

\section{Benchmark Bounds from Direct Tidal Measurements}
\label{Sect:Constr}
The tidal contributions induced by torsion and nonmetricity can, in principle, be constrained by direct measurements of the tidal tensor. A natural class of experiments is provided by gravity gradiometers, such as GOCE, which measure the components of the tidal tensor $\mathcal{T}^i{}_{j}$. The instrument consists of accelerometer
pairs arranged along three mutually orthogonal axes \cite{Rummel2011GOCE}. However, such measurements access only the total tidal tensor. One must therefore still disentangle the Newtonian, torsional, and nonmetricity-induced contributions.

A single gradiometer operating with only one class of
probe matter cannot distinguish between the individual post-Riemannian
contributions in \eqref{tors_tidal_accel} and
\eqref{nonmetr_tidal_accel}. In particular, the irreducible torsion and
nonmetricity sectors produce overlapping mathematical structures in the tidal
acceleration, so that a single-channel measurement constrains only the
corresponding linear combinations of field components. Separating these
contributions would require additional physical input or complementary
measurement channels, for example channels coupling to probes with different
spin or hypermomentum charges. The precise response of such probes, and hence
the mapping between the individual measurement channels and the irreducible
post-Riemannian fields considered here, depends, however, on the
matter-coupling prescription and the underlying MAG model. Since a treatment of such controlled affine-sensitive
probe channels lies beyond the scope of the present analysis, we adopt an
idealized single-field assumption: when deriving a bound for a given irreducible post-Riemannian
sector, only that sector is retained, while all other torsion and nonmetricity
sectors are set to zero. The following estimates should therefore be understood as flat,
componentwise benchmark limits under this idealized assumption, rather than
as separable, independent bounds in a general multi-sector configuration.

\subsection{Derivative Bounds}

The tidal acceleration may be expressed as
\begin{equation}
    a^i
    =
    \mathcal{T}^i{}_{j}V^j,
\end{equation}
where
\begin{equation}
    \mathcal{T}^i{}_{j}
    \coloneqq
    \mathcal{T}_{\rm N}{}^i{}_{j}
    +
    \mathcal{T}_{\rm pR}{}^i{}_{j}
\end{equation}
is the tidal tensor.
Here
\begin{equation}
    \mathcal{T}_{\rm N}{}^i{}_{j}
    \coloneqq
    -\partial_i\partial_j\Phi
\end{equation}
is the Newtonian tidal tensor, while
$\mathcal{T}_{\rm pR}{}^i{}_{j}$ contains the post-Riemannian
contributions induced by torsion and nonmetricity. The corresponding
tidal accelerations are
\begin{equation}
    a^i_{\rm N}
    =
    \mathcal{T}_{\rm N}{}^i{}_{j}V^j,
    \qquad
    a^i_{\rm pR}
    =
    \mathcal{T}_{\rm pR}{}^i{}_{j}V^j.
\end{equation}
If, in a direct tidal measurement, no statistically significant
post-Riemannian residual is observed, and if $\Delta \mathcal{T}$ denotes the uncertainty of the instrument, then the post-Riemannian tidal tensor components satisfy
\begin{equation}
    |\mathcal{T}_{\rm pR}{}^i{}_{j}|
    \leq
    \Delta \mathcal{T}
\end{equation}
for every $i,j = 1,2,3$.
In practice, the instrumental uncertainty can differ between tidal tensor components \cite{Rummel2011GOCE}. For the qualitative benchmark estimates considered here, we approximate these uncertainties by a common effective sensitivity $\Delta\mathcal{T}$.
Under the idealized single-field assumption stated above,
each componentwise uncertainty yields a benchmark bound on the
contribution from the retained irreducible post-Riemannian sector.

More specifically, for an axial-torsion tidal contribution, we obtain
\begin{equation}
\left|
\epsilon\indices{_{ijk}}
\partial\indices{_0}s\indices{^k}
\right|
\leq
\Delta \mathcal{T}
\end{equation}
for every $i,j = 1,2,3$.
The nonvanishing off-diagonal components then imply the componentwise bounds
\begin{equation}
\left|\partial_0 s^i\right|
\leq
\Delta \mathcal{T},
\qquad
i=1,2,3.
\label{bound_ax_t}
\end{equation}
Thus, the time derivative of each spatial component of the axial torsion vector is bounded by the effective tidal tensor sensitivity.
Similarly, for the vectorial-torsion contribution, the componentwise bound yields
\begin{equation}
\left|\partial_j S^i\right|
\leq
\frac{3}{2}\Delta \mathcal{T},
\qquad
i,j=1,2,3.
\end{equation}

Analogous reasoning yields bounds on the irreducible sectors of nonmetricity. The tidal tensor contribution associated with Weyl nonmetricity is given by
\begin{equation}
\frac{1}{2}
\left(
\partial_j W^i
-
\delta^i_j \partial_0 W_0
\right).
\end{equation}
Therefore, the componentwise bound implies
\begin{equation}
\left|
\partial_j W^i
-
\delta^i_j \partial_0 W_0
\right|
\leq
2\Delta\mathcal{T},
\qquad
i,j = 1,2,3.
\end{equation}
Recall that \eqref{kin_consist_irrep} holds along the reference trajectory.
Under the single-field assumption, however, only $W_0$ remains in
\eqref{kin_consist_irrep}, and hence $W_0=0$ along the trajectory. Since
$U^\mu \approx (1,0,0,0)$, it follows that
$\partial_0W_0=0$ along the trajectory to the order considered. Therefore, we find
\begin{equation}
\left|
\partial_j W^i
\right|
\leq
2\Delta\mathcal{T},
\qquad
i,j = 1,2,3.
\end{equation}
For the $\Lambda$ sector, we obtain
\begin{equation}
\left|
\delta^i_j \partial_0 \Lambda_0
-
5\partial_j \Lambda^i
\right|
\leq
8\Delta\mathcal{T},
\qquad
i,j = 1,2,3.
\end{equation}
In the single-field $\Lambda$ case,
\eqref{kin_consist_irrep} implies $\Lambda_0=0$ along the reference
trajectory and hence $\partial_0\Lambda_0=0$ there. Thus,
the bound
reduces to
\begin{equation}
\left|\partial_j\Lambda^i\right|
\leq \frac{8}{5}\Delta\mathcal{T},
\qquad i,j=1,2,3,
\end{equation}
very similar to the Weyl sector.
Finally, for the pseudotensorial sector, the componentwise bound gives
\begin{equation}
\left|
\frac{2}{3}\partial_j d^i
+
\epsilon\indices{_{ijk}}
\partial\indices{_0}b\indices{^k}
-
\frac{1}{3} \partial_0 \Pi_{ij}
\right|
\leq
\Delta\mathcal{T},
\qquad
i,j = 1,2,3.
\end{equation}
Diagonal components constrain the combination $(2/3)\partial_j d^i
-
(1/3)\partial_0 \Pi_{ij}$, while off-diagonal components additionally include the rotational term $\epsilon\indices{_{ijk}}
\partial\indices{_0}b\indices{^k}$.

\subsection{Characteristic Scales}
It is reasonable to ask whether the derivative bounds derived above can be
translated into bounds on the amplitudes of the fields themselves. This
translation is not unique and requires additional assumptions. One may, for
example, integrate the derivative bounds over finite intervals, assume a
Fourier decomposition or a dominant wavelength, use field equations, or
impose functional inequalities under suitable regularity and boundary
conditions. Here we instead adopt a simple characteristic-scale estimate
from scaling analysis. 

For a field $X(x^i,t)$, let $X_{\rm char}$ denote the characteristic size of its variation over spatial scales $L_X^i$ and over a temporal scale $T_X$. We then estimate
\begin{equation}
    |\partial_i X|
    \sim
    \frac{|X_{\rm char}|}{L_X^i},
    \qquad
    |\partial_0 X|
    \sim
    \frac{|X_{\rm char}|}{T_X}.
\end{equation}
These are coarse-grained finite-difference estimates, in the sense that they characterize the typical change of the field over the length and time scales $L_X^i$ and $T_X$, and constrain only the varying part of the field. This is
sufficient here because constant torsion and nonmetricity components do not
contribute to the tidal terms considered.

In the case of axial torsion we have
\begin{equation}
    |\partial_0 s^i|
    \sim
    \frac{|s^i_{\rm char}|}{T_{s^i}},
\end{equation}
and together with \eqref{bound_ax_t} therefore
\begin{equation}
    |s^i_{\rm char}|
    \lesssim
    T_{s^i}\Delta\mathcal{T},
    \qquad i=1,2,3.
\end{equation}
For vectorial torsion, the estimate
\begin{equation}
    |\partial_l S^i|
    \sim
    \frac{|S^i_{\rm char}|}{L_{S^i}^l}
\end{equation}
yields
\begin{equation}
    |S^i_{\rm char}|
    \lesssim
    \frac{3}{2}L_{S^i}^l\Delta\mathcal{T},
    \qquad i,l=1,2,3.
\end{equation}
Taking the strongest directional bound gives
\begin{equation}
    |S^i_{\rm char}|
    \lesssim
    \frac{3}{2}\min_l\{L_{S^i}^l\}\Delta\mathcal{T}.
\end{equation}
Thus, once a given sector is assumed or inferred to vary on specified spatial or temporal scales, the derivative bounds can be translated into benchmark bounds on the corresponding characteristic amplitudes. Fields varying on shorter characteristic spatial or temporal scales are therefore more tightly constrained.

The same reasoning applies to the irreducible nonmetricity sectors. However, in cases where the tidal tensor constrains only a linear combination of derivative terms, no direct bound on the characteristic amplitudes of the individual terms follows without additional assumptions.

\subsection{Rough Estimate}
A fully realistic experimental analysis
would have to take into account the frequency-dependent noise properties of
a concrete instrument. In GOCE-type gradiometry, the instrumental noise is
commonly characterized in the frequency domain by the square root of the
power spectral density with units
$\SI{}{\milli\eotvos}/\mathrm{\sqrt{Hz}}$, that is, millieötvös per square root hertz, where $\SI{1}{\eotvos} = \SI{e-9}{\per\second\squared}$.
A complete signal-level analysis would require specifying the instrument
configuration, noise model and measurement protocol
of a concrete affine-sensitive gradiometric experiment. Since no such device
is currently available, this analysis lies beyond the scope of the present
work. However, we can provide some heuristic estimates based on reported numbers from GOCE.

We employ a representative gravity-gradient noise
level $n_\mathcal{T}\sim \SI{10}{\milli\eotvos}/\mathrm{\sqrt{Hz}}
=\SI{e-11}{\per\second\squared}/\mathrm{\sqrt{Hz}}$ for the more accurate
gradient components, as reported in \cite{Rummel2011GOCE}.
Assuming, for a rough estimate, a flat noise level over the nominal
measurement band $B=[5,100]\SI{}{\milli\hertz}$ \cite{Rummel2011GOCE}, this corresponds to a tidal tensor uncertainty
\begin{equation}
    \Delta \mathcal{T}
    \sim
    n_\mathcal{T} \sqrt{\Delta f}
    \sim \SI{3e-12}{\per\second\squared},
\end{equation}
where $\Delta f = \SI{95}{\milli\hertz}$.
Hence, the bounds derived above give, for example, the axial torsion derivative bound
\begin{equation}
    c^2\left|\partial_0 s^i\right|
    \lesssim
    \SI{3e-12}{\per\second\squared},
    \qquad
    i=1,2,3.
\end{equation}
Here we have restored the factors of $c$ required
for comparison with SI units. Since both $s^i$ and $\partial_0$ have dimension $L^{-1}$, the combination
$\partial_0s^i$ has dimension $L^{-2}$. Multiplication by $c^2$
therefore converts it to $T^{-2}$, the SI dimension of a tidal-tensor
component. Since $x^0=ct$, we have
$\partial_0=c^{-1}\partial_t$ and hence one may equivalently express
$c^2\partial_0s^i=c\,\partial_ts^i$.
Analogous derivative bounds follow for the remaining irreducible sectors and can be translated into characteristic-amplitude bounds once characteristic variation scales are specified.

We stress, however, that these are only heuristic bounds based on GOCE
performance numbers. Since the instrumental noise depends on frequency, the
estimate applies most directly to post-Riemannian tidal signals whose
time variation lies within the sensitive GOCE measurement band. Signals that
vary much more slowly or much more rapidly would generally be constrained
less strongly, in an instrument-dependent way.

\section{Conclusion and Discussion}
\label{Sect:Disc}
In this work we derived the projected deviation equation in general
MAG with torsion and nonmetricity, and analyzed its
weak-field, nonrelativistic limit under the autoparallel hypothesis.
Within this setting, the resulting tidal force law consists, at leading
order, of the Newtonian contribution together with linear corrections
induced by the post-Riemannian sectors. We further classified the kinematical signatures of the irreducible components of torsion and
nonmetricity and, under the single-field assumption adopted here,
derived possible benchmark bounds on the derivatives and characteristic sizes of these fields.

In the Riemannian limit
$S^\lambda{}_{\mu\nu}=0=Q_{\lambda\mu\nu}$, our general projected
deviation equation~\eqref{dev_eq} reduces to the
standard projected deviation equation of Riemannian
geometry as in~\cite{Hawking73}. For a geodesic congruence, this becomes the familiar geodesic deviation
equation of GR. At the weak-field, nonrelativistic level, the same
limit removes all explicit post-Riemannian contributions from the
tidal acceleration law~\eqref{tidal_acceleration_law}, yielding the
standard Newtonian relation
$a^i=-V^j\partial_i\partial_j\Phi$. Thus, within the
affine-sensitive motion considered here, the additional tidal
contributions originate entirely from torsion and nonmetricity.

Our approach differs from earlier studies as follows. In $f(R)$ gravity as studied in~\cite{delaCruz13}, torsion and
nonmetricity vanish and particles follow geodesics. Its
timelike kinematics is therefore recovered from our deviation equation
by setting $S^\lambda{}_{\mu\nu}= 0=Q_{\lambda\mu\nu}$ and $\hat{A}^\mu = 0$. The $f(R)$ corrections enter indirectly
through the field equations used to evaluate the Riemannian curvature
on an FLRW background, and~\cite{delaCruz13} focuses on
cosmological deviation rather than a local weak-field tidal law.
Similarly,~\cite{Harko12} assumes Riemannian geometry, but
curvature-matter coupling generates non-geodesic motion with
$\hat{A}^\mu=f^\mu$.
Reproducing their force-dependent tidal terms therefore would in particular require retaining
$\hat{A}^\mu=f^\mu$ before taking the weak-field limit. Cosmological studies of $f(T)$ and $f(Q)$ gravity
have likewise employed the Levi-Civita geodesic deviation equation,
with the theory-specific field equations determining the
Levi-Civita curvature and background evolution~\cite{Darabi14,Beh21}.
This is appropriate for minimally coupled structureless matter
following metric geodesics, as explicitly emphasized for $f(Q)$
gravity in~\cite{Capozziello24}. In these treatments, torsion or
nonmetricity therefore modifies the deviation only indirectly through
the background dynamics. By contrast, for the affine-sensitive motion
considered here, both fields enter the kinematical deviation law
directly. Weak-field tidal effects have also been studied in
torsion-free Weyl-type $f(Q,T)$ gravity, where $T$ denotes the trace
of the energy-momentum tensor and nonmetricity is restricted to its
Weyl-vector form~\cite{Yang21}. The nonmetricity-matter coupling
produces an extra force $\hat{A}^\mu=f^\mu$. However, since that
analysis employs an unprojected deviation equation, a direct
term-by-term comparison with our result is not immediate. Finally,~\cite{Puetzfeld18} is restricted to
Riemann-Cartan geometry and defines the finite separation between
neighboring worldlines using an auxiliary autoparallel connector,
whereas our infinitesimal congruence construction allows arbitrary
transverse curves. Reference~\cite{Agashe23} treats both torsion and
nonmetricity but derives an unprojected deviation equation rather
than the transverse relative acceleration considered here.

The present analysis should be understood as an idealized
affine-sensitive probe model. More precisely, we studied the projected
tidal response of neighboring timelike curves under the assumption
that the reference motion is governed by autoparallels of the full
affine connection. In this sense, the results derived here isolate the leading
tidal signatures that arise if free fall is sensitive to the full
metric-affine structure. Whether torsion and nonmetricity actually
enter the motion of physical test bodies is, however, a separate
dynamical question, to be decided by the underlying matter couplings
and the corresponding equations of motion. A natural continuation of
the present work is therefore to relax the autoparallel assumption and
study the projected deviation equation for more general test-body
dynamics in metric-affine spacetime, such as those discussed in
\cite{Iosifidis23}.

Although we concentrated on direct tidal
constraints, indirect bounds derived from integrated orbital
observables are equally important and complementary.
Post-Riemannian contributions that induce persistent
rotational or directional distortions act as small perturbing
forces in orbital dynamics and can accumulate over long
time scales.
If their orbit-averaged contribution does not vanish,
they may generate secular drifts of orbital elements such as
the ascending node or periapsis.
Rotational contributions naturally lead to precession-type signatures
that build up over many orbital periods.
In this context, the following observational channels are
particularly natural candidates for further investigation:

\begin{itemize}
    \item \textbf{Gyroscope spin precession:}
    Tidal corrections enter the transport law of spin vectors.
    Vorticity-type contributions may therefore induce additional
    spin precession beyond the standard relativistic effects.
    Integrated along an orbit, even small corrections can accumulate
    into measurable shifts in the precession rate, as in the type of
    observables targeted by Gravity Probe B (however, notice also \cite{PuetzfeldProbing07}) and by frame-dragging
    measurements with laser-ranged satellites such as LAGEOS and
    LARES \cite{Ciufolini16}.

    \item \textbf{Satellite-to-satellite tracking:}
    Missions that continuously monitor the relative motion of two nearby
    satellites effectively probe the evolution of the deviation vector
    between their worldlines.
    Persistent post-Riemannian corrections would therefore appear as
    small perturbations in the inter-satellite dynamics, potentially
    generating characteristic along-track or cross-track signatures as
    well as secular drifts. This makes GRACE and GRACE-FO particularly
    natural reference systems for such effects. In this regard, GRACE-C, the continuation to be launched in 2028, could prove fruitful.

    \item \textbf{Relativistic binary pulsars:}
    In compact binaries, small post-Riemannian corrections can accumulate
    over many orbital periods and manifest themselves through secular
    precession of orbital elements. In particular, the double pulsar and
    related systems provide exceptionally clean timing laboratories for
    spin-orbit and gravitomagnetic effects, including Lense-Thirring-type
    precession signatures \cite{Hu24}.
\end{itemize}

Although such indirect bounds are typically orbit-averaged,
they can yield extremely strong constraints owing to the
remarkable precision and long integration times available
in modern experiments.
A systematic exploration of these integrated observables
within the present framework therefore constitutes a
natural and promising avenue for tightening phenomenological
bounds on torsion and nonmetricity.

Beyond phenomenological considerations, a natural next
step is to extend the present analysis beyond the weak-field
approximation to a fully relativistic treatment.
In strongly curved spacetimes, tidal dynamics plays a central role in the physics of compact
objects.
Since torsion and nonmetricity enter directly into the
deviation equation, they may modify relativistic
precession effects, tidal disruption thresholds, and the
dynamics of strongly gravitating two-body systems.
In particular, in the vicinity of neutron stars or black
holes, post-Riemannian corrections could alter the detailed
structure of tidal stretching and compression, potentially
affecting orbital motion, accretion phenomena, or even
the gravitational-wave signal emitted during inspiral.
A systematic extension of the present framework to the
fully relativistic regime would therefore provide a
direct and conceptually clear route toward identifying
possible strong-field signatures of MAG.

In conclusion, tidal accelerations offer a geometrically
transparent window into
spacetime structure beyond the Riemannian framework.

\section*{Acknowledgments}
The authors thank Mercè Guerrero and the CCGG group for valuable discussions. They further acknowledge the support of the Walter Greiner-Gesellschaft zur Förderung der physikalischen Grundlagenforschung e.V. (WGG), Frankfurt. A.vdV. gratefully acknowledges funding from the Deutsche Forschungsgemeinschaft (DFG, German Research Foundation), Project No.~570900169. D.V. gratefully acknowledges support from the Fueck-Stiftung. M.N.M. gratefully acknowledges support from the Margarete und Herbert Puschmann-Stiftung.

\bibliography{biblio}

\end{document}